# Spotlight on nerves: Portable multispectral optoacoustic imaging of peripheral nerve vascularization and morphology

Dominik Jüstel[1,2,3,†,*], Hedwig Irl[4,†], Florian Hinterwimmer[5], Christoph Dehner[1,3], Walter Simson[6], Nassir Navab[6,7], Gerhard Schneider[4], Vasilis Ntziachristos[1,3,7]

[1] *Institute of Biological and Medical Imaging, Helmholtz Zentrum München, Neuherberg, Germany*
[2] *Institute of Computational Biology, Helmholtz Zentrum München, Neuherberg, Germany*
[3] *Chair of Biological Imaging at the Central Institute for Translational Cancer Research (TranslaTUM), School of Medicine, Technical University of Munich, Germany*
[4] *Department of Anesthesiology and Intensive Care, School of Medicine, Klinikum Rechts Der Isar, Technical University of Munich, Munich, Germany,*
[5] *Department of Orthopaedics and Sport Orthopaedics, School of Medicine, Klinikum Rechts Der Isar, Technical University of Munich, Munich, Germany,*
[6] *Chair for Computer Aided Medical Procedures and Augmented Reality, Technical University of Munich, Germany,*
[7] *Munich Institute of Robotics and Machine Intelligence, Technical University of Munich, Germany*

[†] *These authors contributed equally to this work.*
[*] *Corresponding author. Email: dominik.juestel@helmholtz-muenchen.de*

## Abstract

Various morphological and functional parameters of peripheral nerves and their vascular supply are indicative of pathological changes due to injury or disease. Based on recent improvements in optoacoustic image quality, we explore the ability of multispectral optoacoustic tomography, in tandem with ultrasound imaging (OPUS), to investigate the vascular environment and morphology of peripheral nerves in vivo in a pilot study on healthy volunteers. We showcase the unique ability of optoacoustic imaging to visualize the vasa nervorum by observing intraneurial vessels in healthy nerves in vivo for the first time. In addition, we demonstrate that the label-free spectral optoacoustic contrast of the perfused connective tissue of peripheral nerves can be linked to the endogenous contrast of haemoglobin and collagen. We introduce metrics to analyze the composition of tissue based on its optoacoustic contrast and show that the high-resolution spectral contrast reveals specific differences between nervous tissue and reference tissue in the nerve's surrounding. We discuss how this showcased extraction of peripheral nerve characteristics using multispectral optoacoustic and ultrasound imaging can offer new insights into the pathophysiology of nerve damage and neuropathies, for example, in the context of diabetes.

## Introduction

Localization of peripheral nerves and investigation of their morphology and function is important in clinical medicine and research, for example, to identify nerves for regional anesthesia or during surgery, and for studying the mechanisms underlying peripheral neuropathies. In current clinical routine, peripheral nerves are dynamically localized with ultrasonography, based on anatomical landmarks along the nerve's progression and via the honeycomb-like appearance of the fascicular structure of nerves (*1, 2*). Although ultrasound-guided regional anesthesia has been shown to improve the success rate and safety of the procedure (*2*), electrical nerve stimulation is sometimes needed for verification due to a lack of specificity of ultrasound contrast. US is also employed for visualizing morphological alterations associated with disease and injury (*3, 4*), in particular for diagnosing neuropathies that correlate with a change in the neural cross-sectional area (*4*). However, most acquired axonal neuropathies, including diabetic neuropathy, are not associated with specific morphological changes detectable with US, e.g., a significant nerve enlargement (*4*). The functional abilities of US imaging are limited to visualizing intraneurial blood flow using the Doppler function, which is only sensitive enough to detect substantially increased blood flow in progressed stages of neuropathy (*3*). Other radiological methods such as magnetic resonance imaging (MRI) or positron emission tomography (PET) can identify structural damage and edema (*5, 6*) or regions of high metabolic activity (*5*), but are expensive, not ubiquitously available (*7*), not suited for all patient collectives (e.g., PET – pregnant , MRI – pacemaker), and cannot visualize the vasa nervorum (*6*). Invasive approaches have also been considered for functional nerve investigations (e.g. endoscopic fluorescence-lifetime (sodium fluorescein) spectrophotometry of blood oxygenation in the nerve (*8*), microscopic assessment of biopsies of the sural nerve, or minimally invasive skin biopsies addressing distal small fiber abnormalities (*9*)), however they are only suitable for specific neuropathies, cannot be used for monitoring purposes, and pose a greater risk to the patient.

Vascularization and oxygenation would be valuable features for clinical assessment of peripheral nerve function. For example, diabetic polyneuropathy can cause intraneurial hypoxia by altering the regional blood flow (*10*). Moreover, patients with diabetes (*11*) and peripheral demyelinating diseases (*12*) exhibit changes in their neural water and fat content. However existing radiological modalities have insufficient capabilities to image these parameters, in particular in relation to portable usage. This limitation explains why physical examinations and electrophysiology remain the gold standard for clinical diagnosis of peripheral neuropathy.

Optoacoustic (OA, also photoacoustic) imaging has been proposed as a non-invasive modality that complements existing peripheral nerve imaging approaches by providing optical contrast in tissue dynamically and without harmful radiation exposure. Pilot studies in animal models have demonstrated the feasibility of resolving nerve and vascular features using OA microscopy or macroscopy (*13-16*). A single wavelength pilot study in humans also showcased the ability of OA imaging to resolve the vasculature around nerves. Nevertheless, no systematic study to our knowledge has investigated the performance of OA imaging in human nerves in vivo. Critically, no demonstration or analysis has built on the ability of multispectral optoacoustic tomography (MSOT) to identify endogenous chromophores (i.e., oxygenated and deoxygenated hemoglobin, lipids or water) in nerves by resolving their absorption spectra. MSOT has already proven its value in clinical studies by assessing the vasculature and vascular dynamics in tissues, and visualizing small blood vessels with a performance superior to clinical US (*17-19*). MSOT also resolves tissue oxygenation and metabolic parameters (*20-25*) without the use of contrast agents (label-free operation) or ionizing radiation. Such capabilities would make MSOT a valuable addition to existing nerve imaging modalities by improving contrast for nerve localization and providing new biomarkers for the assessment of nerve health, for example, in the context of diabetes. However, applying MSOT to nerve imaging requires suitable data processing and analysis to overcome the detrimental effects of light absorption in superficial tissue layers and electrical noise in the imaging system.

Recent enhancements in OA image quality (*26-29*) allow MSOT to obtain high-resolution spectral OA data that potentially yields more reliable spectral statistics. For example, correlations between spectral components in regions of interest (ROIs) may detect changes in the relative proportions of tissue components. Based on this development, we herein explore the performance of MSOT for imaging human peripheral nerves in vivo. We performed a pilot study in which three major distal nerves of the brachial plexus are imaged with a handheld optoacoustic-ultrasound (OPUS) system in the upper arm of healthy volunteers. Using the spatial and spectral dimensions of MSOT, we explore its performance in revealing the vasa nervorum and the internal structure of peripheral nerves. Of particular interest was assessing MSOT's ability to observe small intraneurial vessels in healthy peripheral nerves, which has previously not been possible with a handheld system in vivo. We show that data-driven spectral unmixing methods afford a detailed picture of the vasa nervorum, including relative blood oxygenation and the internal structure of peripheral nerves. Individual fascicles can be identified via MSOT contrast of the nerve's connective tissue and its vascular supply. This level of contrast and resolution in multispectral OA images has not been demonstrated previously.

In addition, we highlight the ability of MSOT's high-resolution spectral contrast to sense differences in tissue composition. We introduce metrics to quantify tissue specific features related to the mixing behavior of different spectral unmixing components and demonstrate that these metrics capture the specific spectral contrast of nervous tissue and link it to blood, lipid, and collagen contrast. A refinement of the analysis that investigates clusters in the data associated to specific tissue contrasts reveals fine nuances in spectral contrast, which are potentially sensitive to pathological changes in the substructures of nerves. We discuss how OPUS has demonstrated the most detailed anatomical and functional visualization of nerves achieved so far and elaborate on applications that could give access to early features of neuropathy, potentially leading to earlier diagnosis and deeper insights into pathophysiology.

## Results

### OPUS imaging of peripheral nerves

Fig. 1 illustrates the data acquisition, the anatomy of the imaged region, the morphology of peripheral nerves, and representative US images of the three targeted nerves. Fig. 1A shows the OPUS system used in this study, an iThera Medical Acuity Echo prototype, and the approximate region at the upper limb (red dashed line), a few centimetres proximal of the elbow, where OPUS images were acquired of the three major peripheral nerves in the arm: the ulnar, median, and radial nerve. Each nerve was imaged in two different locations on each arm of the 12 volunteers, resulting in a dataset of 135 images (45 ulnar, 47 median, 43 radial). A qualitative sketch of the cross-sectional anatomy in the imaged region of the arm is given in Fig. 1B, with grey rectangles outlining representative fields of view for the three nerves. Fig. 1C shows the anatomy and vascular environment of peripheral nerves: nerve fibers are sheathed in lipid-rich myelin and clustered in fascicles. The nerve's substructures are covered in collagen-rich connective tissue: the epineurium wraps the nerve, the perineurium wraps the fascicles, and the endoneurium wraps single nerve fibres. Accordingly, the supplying blood vessels are classified by their location as epineurial, perineurial, and endoneurial vessels.

Fig. 1D-F show representative US images of the ulnar, median, and radial nerves, respectively, which are highlighted in purple. The depth of the imaged nerves' centers was $7.10 \pm 3.40$ mm for the ulnar nerves, $8.93 \pm 2.87$ mm for the median nerves, and $14.93 \pm 3.50$ mm for the radial nerves. The green shaded areas in Fig. 1 D-F visualize the variance of the nerve locations in the dataset. These areas also serve as references for the spectral contrast of the nerve tissue. Within these regions, we can identify the typical anatomical environments of the three nerves. The ulnar nerve is superficially located next to the triceps with only smaller vessels (e.g., the ulnar collateral arteries) in its immediate vicinity. The median nerve is close to the brachial artery in a region with several large vessels (e.g., brachial and basilic vein). The radial nerve lies significantly deeper, usually between the brachialis and brachioradialis muscles.

Due to their different environments, the three nerves are not equally well-suited for OPUS imaging. The ulnar nerve is the most accessible due to its superficial location and the absence of strong absorbers in its vicinity. The big vessels in the median nerve's vicinity dominate the contrast and can make it difficult to assess the nerve's OA contrast if, for example, the nerve is below a vessel or if imaging artefacts extend into the nerve region. In addition, the pulsation of an artery close to the median nerve might create motion artefacts that reduce image quality. The radial nerve lies deep in muscle tissue, which absorbs much light and leads to a very low OA contrast of the nerve and its immediate surrounding.

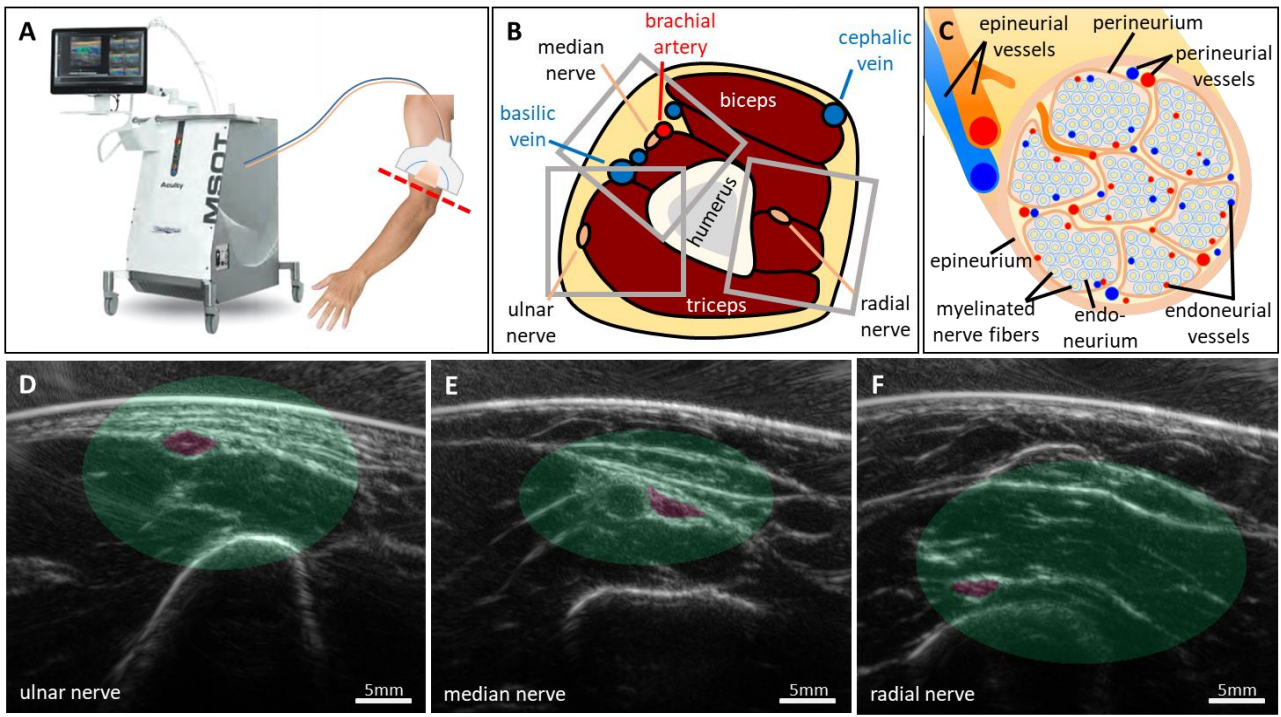

**Figure 1. Data acquisition and anatomy.** (**A**) Co-registered ultrasound and OA images of the three major peripheral nerves of the upper limb were acquired proximal of the elbow (red dashed line) with an iThera medical Acuity Echo prototype. (**B**) Schematic of the cross-sectional anatomy of the upper arm in the imaged region (red dashed line in A). The approximate locations of the fields of view for the three nerves are outlined with grey rectangles. (**C**) Schematic morphology of peripheral nerves. The nerve is organized in fascicles that contain the nerve fibers. The vascular supply is classified according to its location into epineurial, perineurial, and endoneurial vessels. (**D**)-(**F**) Three representative ultrasound images of the three different nerves and their environments. The nerves are highlighted in purple. The areas shaded in green visualize the variance of the nerve locations in the whole dataset.

**Data-driven spectral unmixing reveals tissue specific optoacoustic contrast**

We performed a data-driven spectral unmixing of the MSOT data in order to extract the tissue specific contrast provided by the multispectral images (see Methods section C.2). In contrast to simple linear unmixing, a data-driven unmixing method can capture spectral variations due to wavelength-dependent fluence attenuation. Since nerves are strongly vascularized and contain lipid-rich myelin and collagen-rich connective tissue (Fig. 1C), their spectral OA contrast is expected to be a mixture of the absorption spectra of hemoglobin, water, lipid and collagen.

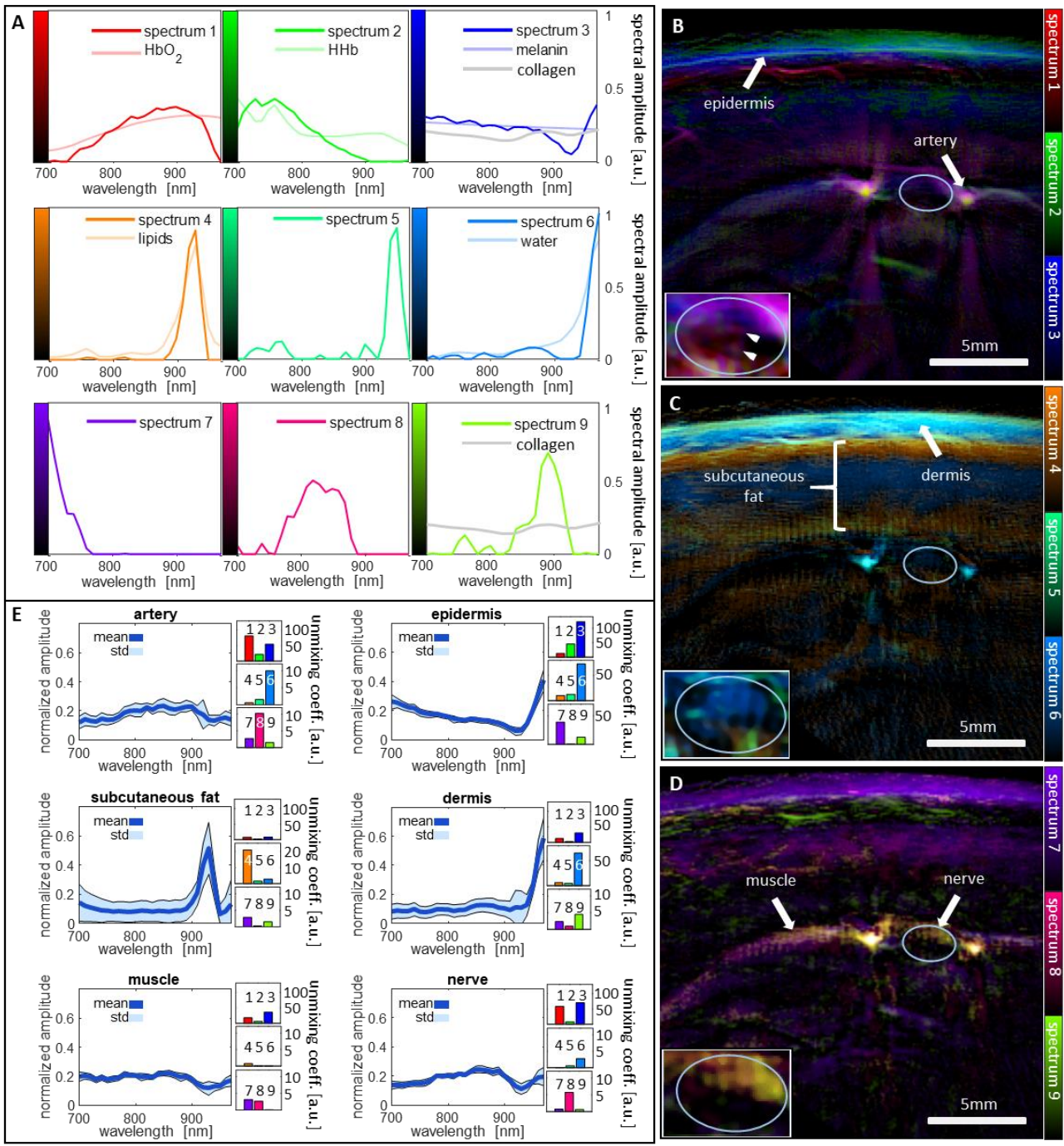

**Figure 2. Spectral unmixing of the MSOT data.** (**A**) The spectral unmixing algorithm found nine distinct spectral components that agree with features of endogenous chromophores (hemoglobin, lipids, water, melanin, collagen). (**B-D**) Spectral contrast of a representative ulnar nerve, visualized by color coding three components per image. (**E**) Mean and standard deviations of the normalized spectra of the structures that are highlighted with white arrows in (C), (D), and (E) – the artery to the right of the nerve, the epidermis, the subcutaneous fat, the dermis, the muscle, and the ulnar nerve. To the right of the graphs, the corresponding mean unmixing coefficients are displayed. HbO2: oxygenated hemoglobin, HHb: deoxygenated hemoglobin.

Fig. 2 shows that the data-driven spectral unmixing successfully captures the specific contrast of endogenous chromophores in the imaged tissues and, contrary to simple linear unmixing, can extract small spectral variations that are hidden behind the dominant absorptions of melanin, hemoglobin, lipids, and water. The qualitative accuracy of the unmixing is validated in a representative image of an ulnar nerve. Fig. 2A shows the nine fundamental spectral components that the unmixing algorithm extracted from the multispectral OA image data, i.e., every measured spectrum is represented as a weighted sum of these components. Four of these spectra closely resemble the absorption spectra of prominent endogenous chromophores: oxygenated and deoxygenated hemoglobin (spectra 1 and 2, respectively), lipids (spectrum 4), and water (spectrum 6). Spectrum 3 has the common slope of the spectra of melanin and collagen, but differs from both in the regions of lipid and water absorption. Instead, the weak absorption peak of collagen around 900 nm is captured in spectrum 9. The remaining three spectral components (spectra 5, 7, and 8) capture variations of the spectra, presumably due to the light absorbed in superficial layers.

Fig. 2B-D visualize the contrast of the nine spectral components in a representative image of an ulnar nerve and its environment by color-coding three of the components in each image. The location of the nerve is indicated by a white ellipse. The inlays show an enlargement of the nerve with adjusted dynamic range to reveal the intraneurial contrast that is otherwise hidden next to the strong signal of blood vessels. The spectral contrast in Fig. 2B highlights the melanin in the skin line, the blood vessels in the skin, and two blood vessels to both sides of the nerve (presumably the ulnar collateral arteries). The strong optoacoustic contrast of these tissues is in agreement with the fact that blood and melanin are the strongest absorbers in tissue. Fig. 2C shows strong contrast from spectra 4-6 in the superficial layers, highlighting the dermis and the subcutaneous fat. The fact that the subcutaneous fat extends until the depth of the nerve is not correctly reconstructed from the OA signal data due to filtering of the acquired signals to suppress low frequency noise and to achieve a better contrast of smaller structures, like nerves and vessels. The perfused tissues below – muscle and nerve – have good contrast in Fig. 2D.

To qualitatively validate the accuracy of the unmixing results, Fig. 2E shows the normalized spectra and standard deviations of six tissues that can be identified in the multispectral images together with their decomposition into the nine spectral unmixing components. The corresponding tissue segmentations are shown in Supplementary Figure 1. The unmixing correctly captures the contrast of the oxygenated hemoglobin in the artery with spectrum 1, the contrast of water and melanin (epidermal melanocytes) in the epidermis with spectra 3 and 6, and the lipid contrast of the subcutaneous fat with spectrum 4. Since deoxyhemoglobin is present in healthy tissue only in mixtures dominated by oxyhemoglobin, and collagen has a low absorption in general, these two chromophores are difficult to unmix. The additional fact that the absorption spectra of these chromophores lack unique features in the considered wavelength range leads to ambiguities in the unmixing. As a consequence, spectra 2 and 3, which both have a slope similar to the spectra of deoxyhemoglobin, collagen, and melanin, can be used to explain spectral data related to any of these chromophores. Most notably, spectrum 3 is often present as a component of blood spectra, for example, in the spectrum of the artery. The strong contribution of spectrum 9 to the spectrum of the dermis indicates light absorption by dermal collagen. The spectra of nerve and muscle are similar to the spectrum of oxyhemoglobin due to the fact that both tissues are well perfused with blood. The dip in the wavelength range of lipid absorption (900-940 nm) indicates the light attenuation due to the subcutaneous fat layer above these tissues.

The inlays in Fig.2 B-D show the intraneurial spectral contrast with two small absorbers (marked by arrows) in Fig.2 B that indicate the presence of two small intraneurial blood vessels. While there is a clear water contrast visible inside the nerve in Fig.2 C, the expected lipid contrast of myelin is not visible. A possible reason for this is the attenuation of light in the respective wavelengths by the subcutaneous fat. The nerve has good contrast in the inlay in Fig. 2D, showcasing the ability of the data-driven unmixing to represent spectral data in deep tissue by capturing the spectral variations due to light attenuation, which are encoded in spectra 7-9.

**OPUS visualizes the vasa nervorum in unprecedented detail**

A direct application of the spectral unmixing is the visualization of the vasa nervorum which play a central role in several neurological disorders. To this end, Fig. 3 shows that OPUS can visualize this vasculature with very good contrast, exceeding the capabilities of other in vivo nerve imaging modalities. All panels of Fig. 3 show images of US contrast (gray) overlaid with two OA channels related to hemoglobin contrast (spectrum 1 – oxyhemoglobin in red, spectrum 2 – deoxyhemoglobin in blue) for six different peripheral nerves.

Fig. 3A-B demonstrate the high MSOT contrast of blood vessels running along and inside of two superficial ulnar nerves. In Fig. 3A, three blood vessels are clearly visible in the MSOT image to the right of the ulnar nerve. While the middle of these vessels, which is the ulnar collateral artery, could also be identified via its pulsation in the US images or via Doppler, the very small epineurial vessels to the left of the nerve and the intraneurial vessels, which are clearly visible in the MSOT images (marked by arrows), cannot be visualized with US. Fig. 3C-D show two ulnar nerves with supplying vessels branching into the nerves visible in the MSOT channels, another detail of the vasa nervorum that cannot be visualized with US. Fig. 3E-F show images of a median and radial nerve, which showcase MSOT's ability to visualize vasculature around and within nerves that are deeper than 1.5 cm and below muscle tissue. The median nerve in Fig. 3E with the brachial artery to its left has clearly visible intraneurial vessels, while the radial nerve in Fig. 3F is accompanied by two medium-sized vessels with clear hemoglobin contrast even though the nerve is below a layer of muscle tissue. A small epineurial vessel is even visible on top of the nerve. Intraneurial blood flow could previously

only be visualized in patients with severe neuropathy (*3*), making the images in Fig. 3 the first visualizations of intraneurial vessels in healthy probands in vivo with a handheld system.

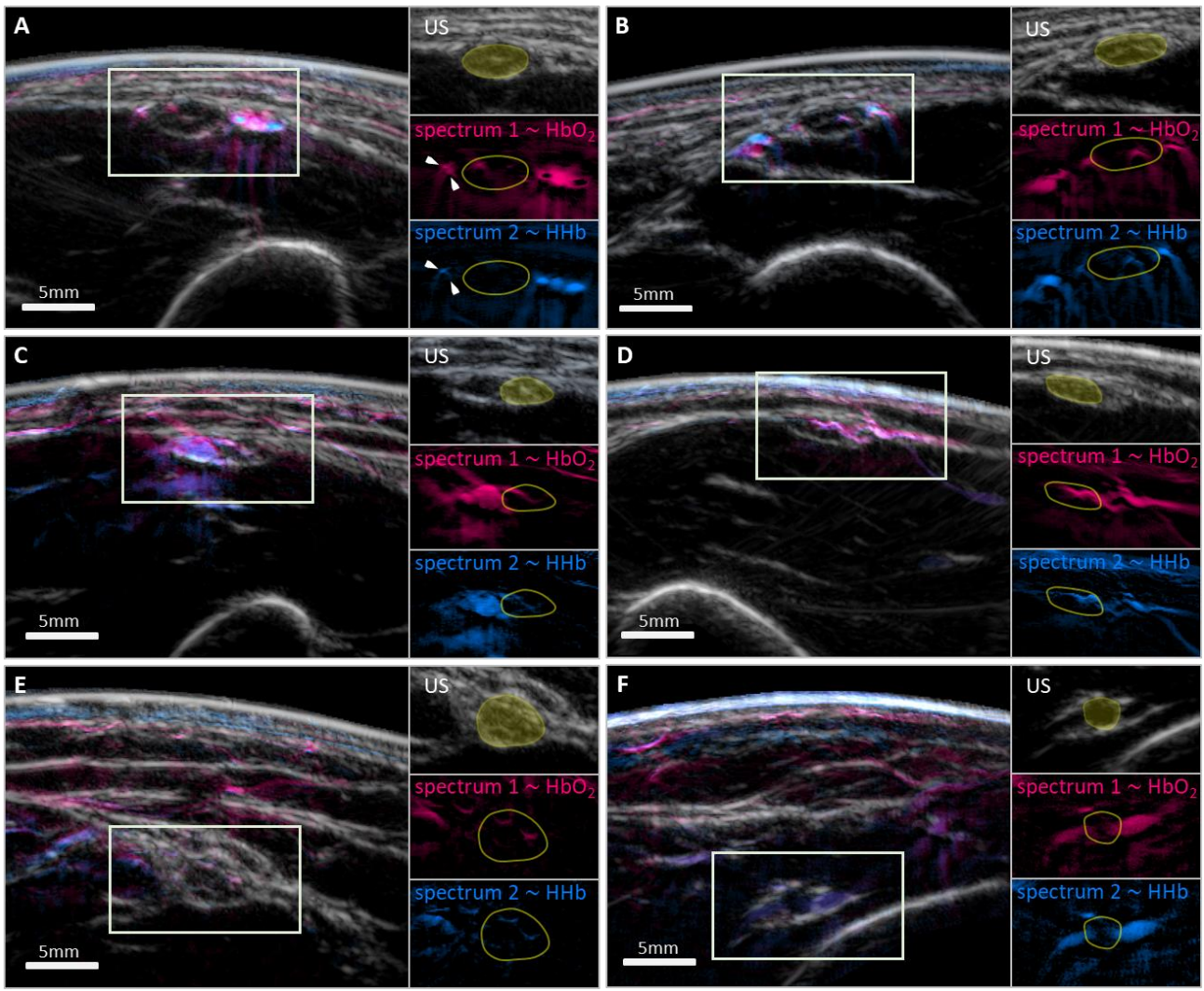

**Figure 3. Visualizing the vasa nervorum.** Overlay images of ultrasound (US, gray) and two spectral OA components associated to oxygenated (spectrum 1, red) and deoxygenated (spectrum 2, blue) hemoglobin. The single components in the regions outlined with rectangles are shown in three small panels next to the images with the nerve highlighted in yellow. (**A**)-(**B**) Two ulnar nerves with intraneurial vessels clearly visible in the OA channels. Arteries (presumably the ulnar collateral arteries) and veins at comparable depths can be spectrally distinguished. (**C**)-(**D**) Two ulnar nerves with vessels branching into the nerve. (**E**) A median nerve with visible intraneurial vessels. (**F**) A radial nerve with two accompanying vessels of medium size, and a small (presumably epineurial) vessel on top.

The images in Fig. 3A-B also show that arteries and veins at comparable depths can be clearly distinguished by their spectral contrast. Specifically, veins have a stronger signal in spectral component 2 due to the higher percentage of deoxygenated hemoglobin. This qualitative distinction is demonstrated close to the system resolution limit of about 200 μm in panel A, where only one of the very small vessels (marked by arrows) is visible in spectral component 2, indicating that it is a vein.

The visualization of the vasa nervorum strongly depends on the depth and environment of the nerve, as shown by the percentages of nerves for which epineurial or intraneurial vessels could be visually identified (Table 1). Vessels are identifiable in two thirds of the imaged ulnar nerves, less than half of the median nerves, and in only a few radial nerves. In addition, a diffuse blood contrast is visible in almost all of the 135 imaged nerves, indicating the contrast of subresolution vasculature, which is a valuable feature for the assessment of nerve perfusion.

**Table 1. Visible vessels of the vasa nervorum.** Numbers of nerves for which individual vessels could be visually identified in MSOT images.

| visible vessels | ulnar nerve (n=45) | median nerve (n=47) | radial nerve (n=43) |
|---|---|---|---|
| epineurial (pct.) | 30 (67%) | 17 (36%) | 3 (7%) |
| intraneurial (pct.) | 16 (36%) | 11 (23%) | 1 (2%) |

**OPUS can access the internal structure of peripheral nerves**

Fig. 4 shows that OPUS can resolve the internal structure of peripheral nerves, i.e., the organization of nerve fibres into fascicles. The honeycomb-like contrast, which can usually also be visualized with high-frequency ultrasound, is visible in MSOT in spectral channels related to blood and collagen contrast. Thus, MSOT can link the structure of nervous tissue to its function.

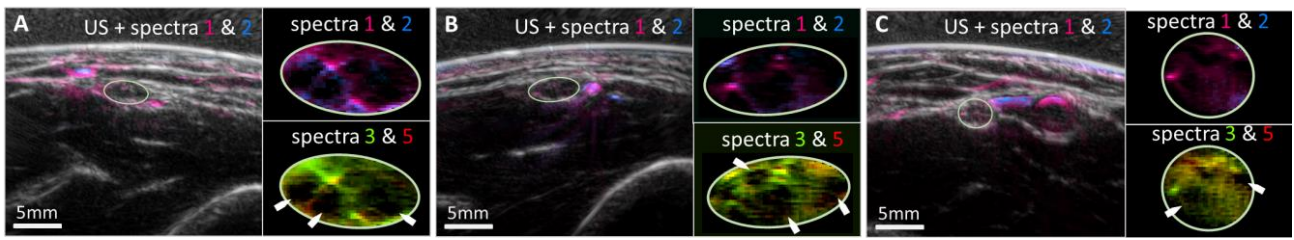

**Figure 4. Visualizing the internal structure of nerves.** The internal organization of nerves into fascicles and their vascular supply are visualized with good contrast by data-driven spectral unmixing. Each panel shows an overlay of US and MSOT hemoglobin contrast (spectra 1 and 2) along with two enlarged images of a nerve with contrast from spectral components 1 and 2 (hemoglobin contrast) and components 3 and 5 (contrast related to collagen and hemoglobin). (**A**) and (**B**) show two ulnar nerves, (**C**) shows a median nerve.

The three panels of Fig. 4 show overlay images of US and vascular contrast of two ulnar (A and B) and a median nerve (C), which are outlined with ellipses. The respective two subpanels display the intraneurial contrast of spectra 1 and 2 (hemoglobin contrast) and spectra 3 and 5 (contrast associated with collagen and hemoglobin). While the upper subpanels only show the intraneurial vessels, fascicles can be clearly identified in the lower subpanels as negative contrast (marked with arrows) outlined by the contrast of the connective tissue and the intraneurial vasculature. The images also again emphasize the intraneurial vascular detail that is accessible with MSOT.

**Correlation metrics reveal differences between the MSOT contrast of nervous and reference tissue**

Going beyond representative images, we performed a correlation analysis on the whole dataset to confirm the observation that OPUS can capture the specific spectral contrast of peripheral nerve tissue. The goal was to understand how different spectral components mix in tissues, and which mixtures are observed in the data. We also sampled reference spectra from the nerves' surroundings to investigate how the spectral contrast of nerves differs from that of other tissues at similar depths. We focus on the ulnar nerve in the rest of the paper because of the superior quality of the spectral data acquired in these more superficial nerves.

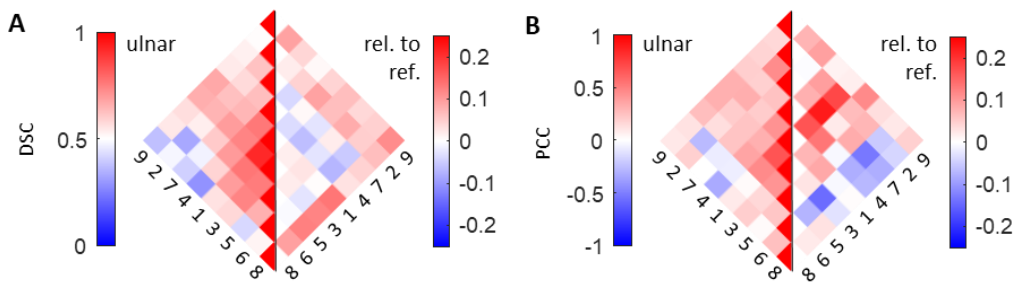

**Figure 5. Correlation analysis of the spectral data acquired from ulnar nerves.** (**A**) Pairwise Sørensen-Dice coefficients (DSC) of the spectral components of the ulnar nerve and their differences from the reference spectra that were sampled from the surrounding tissue. (**B**) Pairwise Pearson correlation coefficients (PCC) of the spectral components of the ulnar nerve and their differences from the reference spectra that were sampled from the surrounding tissue. The spectral components are reordered for better visual interpretability

We investigated the spectral mixtures quantitatively using two correlation metrics – the Sørensen-Dice coefficient (DSC) and the Pearson correlation coefficient (PCC) (see Methods section C.4). In the context of spectral components, the DSC can be interpreted as the tendency of two components to mix: low DSCs indicate that components rarely mix, while high values indicate that components often appear together. The PCC in turn quantifies the type of mixtures: positive values for fixed ratio mixtures, low absolute values for random mixtures, or negative values for competitive mixtures.

Fig. 5 shows the correlation metrics for the ulnar nerve and the reference data; the pairwise DSCs of the spectral components in Fig. 5A and the pairwise PCCs in Fig. 5B. The results for the median and radial nerves can be found in Supplementary Fig. 2. The left half of each panel displays the metrics for the spectra inside the nerve segmentations, while the right side shows the differences between correlations in nervous tissue and reference tissue sampled from the nerve's surrounding, wherein positive and negative values indicate higher and lower correlations inside nerve ROIs, respectively.

The results in Fig. 5 confirm the biological accuracy of the unmixing results by reproducing the expected spectral mixtures in tissue and the expected differences between nervous and reference tissue. In particular, they show that in nervous tissue, relative to reference tissue, all components that relate to features of the collagen spectrum have higher mutual correlations, and that hemoglobin contrast is stronger correlated to lipid and collagen contrast. This observation is in agreement with the expected perfused connective tissue contrast within the nerve.

The left side of Fig. 5A shows that spectra 1 and 3 have the highest DSC with each other, and also high DSC with all other spectra, confirming that they are the main components used for blood contrast, which is the dominant contrast of MSOT. Spectra 4 (lipid peak) and 8 have a very low DSC in general, confirming that spectrum 8 encodes blood contrast in deeper tissue below the subcutaneous fat, and highlighting the ability of data-driven unmixing to capture such spectral coloring effects. The low DSC of spectra 2 and 6 (water peak) shows that spectrum 3 with its small peak at 970 nm encodes the water contrast in mixtures with blood instead of the pure water spectrum 6, demonstrating that the introduced metrics help to interpret unmixing results.

The left side of Fig. 5B shows that the PCC values are generally similar to the DSC values (left side of Fig. 2A), showing that spectral components that often mix do so in stable mixtures, while competitive components rarely mix. The differences between the PCCs of nervous and reference tissue show that spectrum 1 (oxyhemoglobin) has higher correlations with spectra 2-5 and 7, which relate to blood, lipid, and collagen contrast, reproducing the expected contrast of perfused connective tissue and lipid-rich fascicles in the nerve. In addition, spectrum 9 (collagen bump) correlates strongly with all decreasing spectra (collagen slope; spectra 2,3, and 7), which strongly suggests a relation to collagen contrast. The negative correlations of spectra 6 with spectra 2 and 3 shows that intraneurial water contrast in mixture with blood is preferably encoded with spectrum 3.

**Hierarchical clustering finds specific details of nervous tissue in MSOT data**

Fig. 6A-C show the full complexity of the MSOT data with a UMAP embedding of the standardized unmixing data (see Methods section C.5) of the ulnar nervous and reference tissue in three versions colored as in Figure 2B-D. The UMAP algorithm arranges the data in a plane in which every point represents an OA spectrum in the dataset and seeks to keep the spatial and statistical relations between points as close to the original 9-dimensional data as possible. Every cluster in this representation represents the data from one of the $512 = 2^9$ possible mixtures of the nine spectral

unmixing components (Fig. 2A). Due to this intricate structure of the data set, the correlations in Fig. 5B potentially suffer from an averaging effect that cancels out variations in correlations within the subclusters of the data.

We performed a hierarchical clustering of the unmixing data (see Methods section C.5), showing that mixtures of specific spectral unmixing components are consistently used to represent similar spectral contrast, allowing to target specific tissue contrast via clusters in the data. Analyzing the correlations in leafs of the clustering tree, we confirm that the correlations shown in Fig. 5 indeed are affected by averaging effects and show that differences in correlations are specific to leafs of the clustering tree, i.e., specific to the tissue context.

Fig. 6D shows a dendrogram of the clustering tree for the ulnar nerve data. The height of branchings in the tree indicate the dissimilarity of the mean spectral shapes in the two subclusters. We found 409 mixtures to be present in the dataset, with a subcluster of 141 mixtures (34.5% of all leafs) containing most of the contrast. To arrange the 409 mixtures meaningfully, the hierarchical clustering was guided by the mean spectral shape, i.e., the average normalized spectrum. The first two branchings of the tree divide the data coarsely into spectra dominated by fat contrast (yellow shading), spectra that belong to the main blood and perfused tissue contrast (red shading), and spectra of superficial layers with strong water contrast, like skin (green shading).

The relative numbers of pixels contained in each leaf of the clustering tree are plotted along the semicircle delineated by the leafs of the tree, both for the segmented nerve data (red line) and the sampled reference data (blue line). These plots are interpreted as spectral fingerprints of nervous tissue and reference tissue, respectively. Several regions display clear differences. Six interesting clusters, labelled C1-C6, are highlighted by red boxes in the clustering tree: the largest clusters (C3-C6), clusters with clear differences between the ulnar nerve and reference (C1, C3, C4), and specific contrasts (C1, C2; skin and lipid contrast). The six small graphs above the spectral fingerprints show the mean spectral shapes and standard deviations of the six clusters, demonstrating that mixtures are consistently used to represent similar spectral shapes. This implies that specific tissue contrasts can be targeted via hierarchical clustering of spectral OA data. For example, cluster C1 contains specific spectral mixtures of skin and is therefore not expected in nerve tissue, and cluster C2 contains the pure lipid contrast. Clusters C3-C6 represent the main MSOT contrast, containing spectral shapes similar to those of blood vessels, muscle tissue, and nervous tissue (see Fig. 2E).

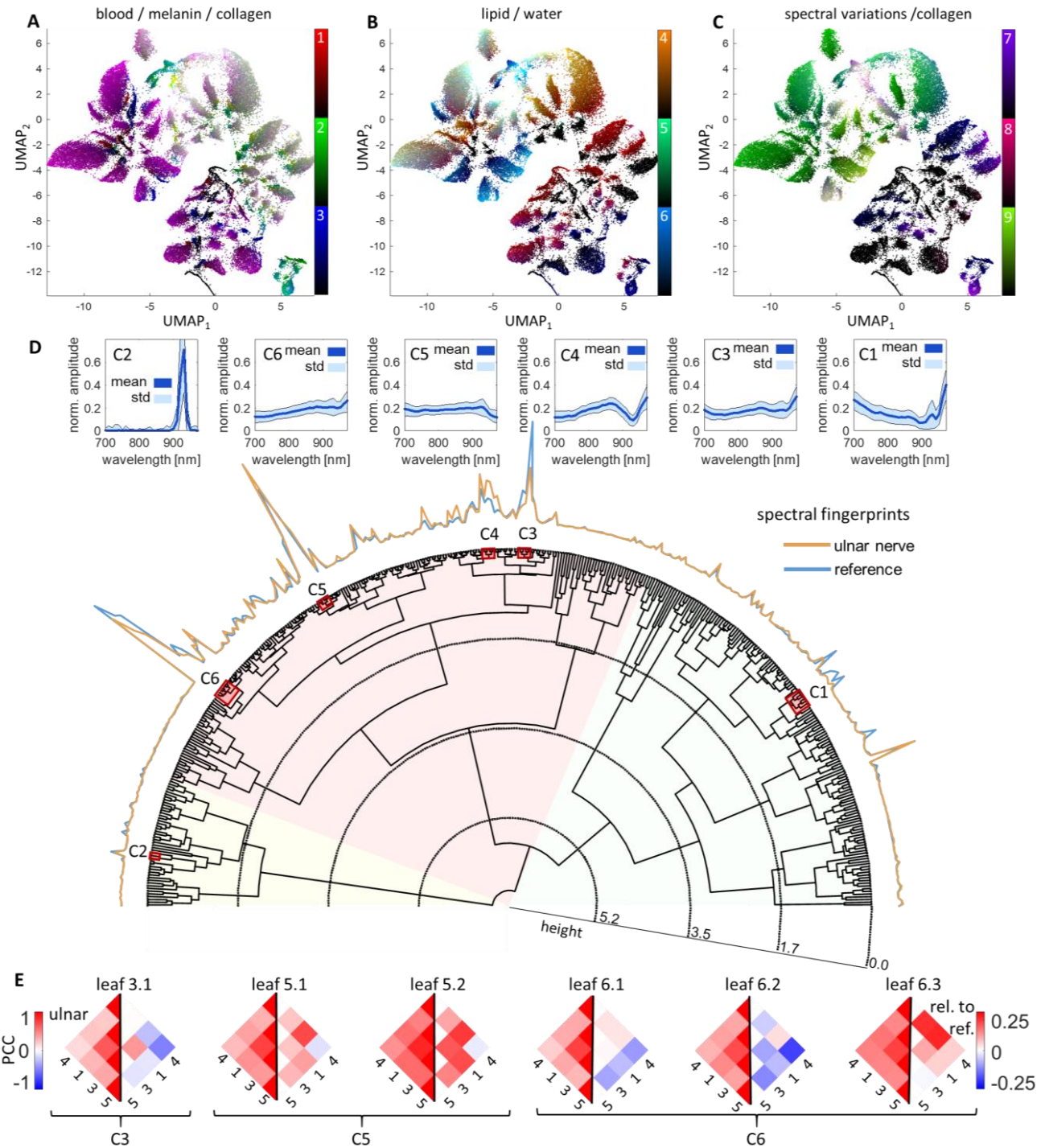

**Figure 6. Hierarchical cluster analysis and UMAP embeddings of data acquired in ulnar nerves.** (**A**)-(**C**) UMAP embeddings of the standardized unmixing coefficients of the ulnar nerve dataset show the complexity of MSOT data. The three versions are colored as in Fig. 2. (**D**) Polar dendrogram of the hierarchical clustering tree. The spectral fingerprints, i.e., the relative number of pixels contained in each leaf, are plotted along the semicircle for the ulnar nerve data (orange) and the reference data (blue). Six clusters, labelled C1-C6 are highlighted in red, with the mean and standard deviations of the spectral shape given in the graphs arranged above the spectral fingerprints. (**E**) Pairwise PCC of the four spectral components present in all the six largest leafs of the clustering tree (called leaf 3.1, 5.1-2, and 6.1-3 according to the clusters they belong to).

Fig. 6E shows the PCCs of the four spectral components that are present in all of the six biggest leafs in the clustering tree (left sides, spectra 1, 3, 4, 5) and the difference between the PCCs of nervous an reference tissue (right sides). The PCC values in nervous tissue consistently agree with the results in Fig. 5B. However, while leafs 5.1, 5.2, and 6.3 show similar correlation differences as in Fig. 5B, the leafs 3.1, 6.1, and 6.2 show significant deviations in PCC. A variation of correlation differences is expected in different substructures. For example, while the perfused connective tissue of the nerve and the fascicles are unique to nervous tissue, blood vessels will have a similar spectral contrast within and outside of nerves. Additional details of the leafs in clusters C3-C6 are shown in Supplementary Figure 3.

**Discussion**

With this pilot study on peripheral nerve imaging using a handheld OPUS system, we demonstrate that hybrid OA and US imaging can access multiple structural and functional parameters of healthy superficial peripheral nerves when combined with a suitable data processing pipeline. The OPUS images of the vasa nervorum (Figs. 3-4) and the connective tissue of peripheral nerves (Fig. 4) achieve a level of detail that has previously not been possible with a handheld OPUS system. Moreover, a dedicated analysis of the data revealed that spectral statistics carry detailed information about the specific spectral contrast of nervous tissue, thereby building a methodological foundation for OA radiomics in peripheral neurology and other medical fields. Since OPUS can be easily integrated into the clinical workflow, the study confirms that this modality has the potential to become a routine tool for the assessment of pathological changes in peripheral nerves.

Peripheral neuropathy, in particular diabetic neuropathy, affects millions of people and can cause serious complications and disabilities (e.g., amputation) (*9, 10*) and increase healthcare costs (more than $10 billion in the United States every year (*30*)). More than 50% of diabetics suffer from polyneuropathy, making it the most prevalent complication of the disease. Screening for neuropathy in diabetics currently does not employ any imaging method, but consists of physical examination and questionnaires, which implies that functional loss is already apparent when symptoms are detected. Despite being a worldwide health issue, the specific mechanisms underlying this polyneuropathy are poorly defined (*30*), though vascular and metabolic factors are thought to play pivotal roles (*10*). As demonstrated in this pilot study, OPUS can assess both vascular and metabolic features of peripheral nerves simultaneously in vivo.

The demonstrated high contrast of the vasa nervorum in MSOT images (see Fig. 3) opens the possibility of investigating pathological changes in the vascular supply of nerves. Indeed, ex vivo and animal studies provide contradictory findings regarding the vasa nervorum of diabetic patients. Endothelial hyperplasia, basement membrane thickening, vascular occlusions, and proliferation seem to play a central role in diabetic microangiopathy and can precede clinical symptoms (*10, 12*). Clinical surveys have also reported reduced perfusion of the diabetes-affected nerve, a change in microcirculation, and endoneurial hypoxia (*10*). Presumably, one of the phenomena contributing to decreased oxygen supply in diabetic neuropathy is arterio-venous shunting. A review by Cameron et al. highlighted correlations between reduced blood flow, intracapillary oxygen saturation, and neurophysiological outcome. Damage in unmyelinated fibers, which regulate the endoneurial blood flow to control arterio-venous shunting, can be seen in very early stages of disease (*10*). With these processes preceding functional loss, improved visualization of epi- or perineurial vessels could give new insights into disease pathogenesis. In this regard, the sensitivity of ultrasound Doppler is insufficient to detect blood flow in healthy nerves (*3*). Increased blood flow in sagittal Doppler imaging has been detected mostly in nerve pathologies and is therefore interpreted as non-physiological (*3, 31*). In our measurements of healthy and young individuals, intraneurial vessels were visualized with good contrast, and veins and arteries could be distinctly identified. Therefore, pathological changes in the nerve vasculature that cause a hypo- or hyperperfusion, e.g., due to arterio-venous shunting, are likely visible in MSOT. In fact, MSOT-based oxygen saturation estimation has already been used to detect vascular malformations (*32*). Thus, the prevailing question of whether hyperperfusion precedes or follows a reduced oxygen supply could potentially be answered with MSOT.

By accessing the water and lipid contrast in peripheral nerves, OPUS imaging can potentially identify several further changes related to diabetic neuropathy. The mild nerve enlargement associated with diabetic neuropathy, for example, might be explained by swelling (*4*). MRI animal studies suggest this is due to an increased intracellular and concomitant decreased extracellular water content caused by augmented amounts of intracellular sorbitol (*11*), an osmotically active product in the polyol pathway. The literature strongly supports that hyperglycaemia-driven excessive metabolism through the polyol pathway is an important pathogenic mechanism of diabetic neuropathy (*12, 33*). Regarding lipid contrast, sural nerve preparations from patients with diabetes show alterations in myelin sheaths up to full segmental demyelination (*12*); theoretically this would appear in MSOT as a decrease in OA contrast in the wavelength range of the lipid peak. With increasing sensitivity in diagnostics, OPUS might further elucidate or even uncover new aspects in pathophysiological processes. It could potentially show abnormal changes that precede functional loss and hence serve as a screening tool and follow up examination on diabetic peripheral polyneuropathy.

As an imaging tool with additional contrasts that can be added to the broadly used ultrasound, OPUS could also help clinical examiners identify nerve tissue. For example, the detection of peripheral nerves or plexi with ultrasound for surgical or invasive treatments (e.g., regional anesthesia) is a common clinical procedure. Even without being the structure of interest, visualization is important to avoid damaging small nerves in the immediate vicinity during invasive procedures like placing a central venous catheter. However, in some cases, nervous tissue cannot be visualized with US because of a variation in anatomy, a small size, or poor contrast. Potential harm can arise due to poor visualization and a subsequent lack of identification (*2*). In this regard, combining the demonstrated ability to extract tissue specific features with machine learning methods for nerve detection and segmentation are a very promising approach.

While spectral unmixing of OA data has the potential to quantitatively determine chromophore concentrations in tissue, several factors limit the feasibility of this task. Spectral coloring due to light fluence attenuation, similarity of absorption spectra of chromophores (e.g., melanin, collagen, and deoxyhemoglobin), and algorithmic biases can lead to biologically implausible unmixing results. In general, the question of whether a contribution of a spectral component truly indicates the presence of the corresponding chromophore is often not addressed in the literature with sufficient detail. A correlation analysis as shown in Fig. 5 can reveal these implausibilities and thereby help to interpret the results, to optimize unmixing algorithms, and most importantly, to prevent incorrect conclusions being drawn from spectral unmixing results.

The introduced hierarchical clustering of the spectral data (Fig. 6) shows that a dedicated data analysis pipeline is necessary to assess the details in spectral contrast. A simple probabilistic model for the unmixing coefficients allowed us to standardize the contrasts of different spectral components and analyze their correlations, even though their contributions to the overall contrast varied by orders of magnitude. The clustering in turn gives access to these correlations on the level of specific tissues. This rich source of information in clinical OPUS datasets is the ideal foundation for OPUS radiomics studies.

In this pilot study we focused on healthy individuals and our findings suggest the possibility to detect morphological and metabolic changes in pathologic nerves. This conjecture needs to be investigated in further studies in patients with pathologically altered nerve anatomy or function. Moreover, the findings need to be correlated against existing tools including electrophysiology, Doppler US, and MRI. A crucial limitation for this application in neuroimaging is the low penetration depth of OA imaging. The best contrast by far was obtained for the very superficial ulnar nerve. While focal neuropathies in nerves that lie deeper, like the proximal ischiadic nerve, can currently not be accessed with OA imaging, further advances in the technology that improve image quality deeper in tissue are anticipated and will expand its applicability.

In summary, we demonstrated that OPUS is a versatile modality for peripheral nerve imaging. With its tissue-specific label-free optical contrast of endogenous chromophores, OPUS could improve nerve localization in difficult situations. Moreover, OPUS' capacity to assess a nerve's lipid, water and collagen contrast and its vascular environment potentially expands the field of peripheral nerve imaging to a broad range of neuropathies by illuminating features of nerve morphology and function that have been hidden so far. Understanding dysfunctional processes is crucial to find powerful solutions in prevention and treatment, while early therapy can change the course of most diseases profoundly.

**Materials and Methods**

**A. Imaging systems**

Two imaging systems were used in this study: a commercial US system for localization of the peripheral nerves and a hybrid OPUS system to acquire the coregistered OA and US data.

The OPUS system is a custom prototype of the Acuity Echo system (iThera Medical GmbH, Munich, Germany); it has a tunable laser that illuminates tissue with laser pulses of ~8 ns duration with an energy of 16 mJ and a repetition rate of 25 Hz, thereby staying within the energy exposure limits defined by the American National Standards Institute. The acoustic part of the system consists of a circularly curved linear US transducer array with a 6 cm radius, 145° angular coverage, and 256 single piezo elements, which were used both for reflection-mode US imaging (~6 MHz excitation frequency) and OA detection (4 MHz central detection frequency). The probe is filled with heavy water as a coupling medium. Acoustic signals are recorded at a sampling frequency of 40 MHz.

Hybrid MSOT and US imaging is realized with a schedule that acquires OA data at a rate of 25 Hz, cycling through 28 different wavelengths (700–970 nm in steps of 10 nm), while US data is acquired in between OA acquisitions at a rate of 6.25 Hz. The US system is operated in a synthetic aperture mode, acquiring reflection data from 256 single transducer transmission events with half the detector array receiving the response. The resulting data consist of multispectral data stacks that are acquired at ~0.89 Hz with 4 US frames acquired per MSOT data stack.

**B. Study protocol**

We investigated the three major distal nerves of the brachial plexus (ulnar, median, and radial) of 12 young healthy volunteers with OPUS. The three nerves were imaged at two different locations of the arm on both sides (= 12 acquisitions per person, 144 total). Informed consent was obtained after the nature and possible consequences of the study were explained. We also collected basic personal parameters (age, sex, height, and weight) and data regarding body composition (BMI, body fat, arm diameter). The latter parameters were recorded because the amount of body fat potentially changes the nerve tissue contrast due to the lipid-rich myelin sheaths of nerve fibers. For examination, the volunteer lay on a bench facing the examiner. A summary of the study cohort is given in Table 2.

**Table 1. Study cohort.** Basic personal and body composition parameters.

| n=12 (8m, 4f) | age [y] | height [m] | weight [kg] | BMI [kg/m$^2$] | bodyfat [%] | arm diam [cm] |
|---|---|---|---|---|---|---|
| *mean ± std* | 29.1 ± 3.2 | 1.79 ± 0.11 | 71.8 ± 16.3 | 22.2 ± 2.9 | 16.0 ± 4.4 | 25.9 ± 2.7 |
| *min – max* | 24 – 36 | 1.60 – 1.93 | 49.0 – 107.0 | 17.6 – 28.7 | 7.0 – 25.0 | 21.0 – 30.0 |

For localizing the nerves, an experienced examiner (anesthetist) screened the arm with a commercial US device. The main peripheral nerves were identified and localized well with US, based on their typical appearance in US images with a reflecting outline and a honeycomb-like contrast within. Anatomical landmarks along the progression of the nerve were used for confirmation. Once the respective nerve was identified, the region of interest was centered and the location marked. We took snapshots of the ultrasound image at that specific site for a first qualitative segmentation of the nerve and to help re-identify it on the Acuity Echo US images.

After locating the nerve with the Acuity Echo US image, we acquired OA data for ~10 sec (~10 multispectral frames). We also acquired the raw US data for the last ten recorded US frames. Because of the different features of the Acuity Echo US and the commercial one, we documented a qualitative segmentation of the nerve in the Acuity US image by photograph.

After cleaning the data (missing data, strong motion, or other artefacts), a total of 135 multispectral frames were available for further analysis. Quantitative segmentation of nerve tissue was performed in the reconstructed Acuity Echo US images by an expert using the qualitative segmentations obtained during the measurements for guidance.

**C. Data analysis**

C.1 Image reconstruction

Both OA and US images were reconstructed in a 4x4 cm field of view at the center of the detector ring and at a resolution of 100 µm, using speed of sound values of 1397 m/s and 1465 m/s for the coupling medium in the detector cavity and the imaged tissue, respectively.

The acquired US data was reconstructed from the synthetic aperture data with a delay-and-sum algorithm after preprocessing. The data was bandpass filtered (3 – 6 MHz) and processed with a spiking deconvolution (*34*). For the delay-and-sum algorithm, the time-of-flight was determined and included the refraction of acoustic waves at the interface between the tissue and detector cavity to achieve a correct coregistration with the OA data. Finally, the delay-and-sum data was integrated via a mixed averaging and maximum intensity approach and logarithmically transformed to obtain the US images.

OA images were obtained from the acquired OA signal data after bandpass filtering (500 kHz - 8 MHz), a deep-learning-based denoising of the sinograms (*28*), cropping the signals according to the field of view, model-based reconstruction with refraction and impulse response correction (*26, 27*) that was regularized with both a Tikhonov regularizer to filter out the noise due to the limited detector coverage, and a Laplacian regularizer to counteract subresolution artefacts. The regularization parameters were determined via the L-curve.

C.2 Data-driven spectral unmixing

To eliminate the influence of artefacts (acoustic reflections below bones, noise above the detector membrane, regions with bad coupling, etc.), regions of interest containing meaningful spectral data were manually segmented, resulting in a dataset of ~12M spectra. To extract the spectral contrast from this dataset, the spectra were blindly unmixed via a non-negative matrix factorization (NMF) (*35*) into nine spectral components, using mixed Frobenius and entrywise $L^1$-regularization. More precisely, arranging the spectra in a non-negative matrix $S \in \mathbb{R}^{N \times 28}$, $S \geq 0$, where N is the number of spectra that are considered and the second dimension contains the values at the 28 different wavelengths, the following optimization problem was solved

$$(W, H) \coloneqq \arg \min_{(W,H) \geq 0} \tfrac{1}{2} \|S - WH\|_F^2 + \lambda_1 (\|W\|_1 + \|H\|_1) + \tfrac{1}{2}\lambda_F (\|W\|_F^2 + \|H\|_F^2),$$

where $\|M\|_F \coloneqq \left(\sum_{i,j} m_{i,j}^2\right)^{1/2}$ and $\|M\|_1 \coloneqq \sum_{i,j} |m_{i,j}|$ denote the Frobenius norm and the entrywise $L^1$-norm of a matrix $M = (m_{i,j})_{i,j}$, respectively, and $M \geq 0$ is an entrywise inequality, meaning that M is a non-negative matix. The matrices $W \in \mathbb{R}^{N \times 9}$ and $H \in \mathbb{R}^{9 \times 28}$ contain the coefficients and spectral components, respectively. The number of components and regularization parameters $\lambda_1 = 80$ and $\lambda_F = 20$ were selected via parameter space exploration and meaningfulness of the resulting spectral components, yielding a relative error $\|S - WH\|_F^2 / \|S\|_F^2$ of 0.72%. The strong entrywise $L^1$-regularization was chosen to promote a maximally sparse decomposition of the spectra, guided by the fact that the spectral contrast of biological tissue is composed by a small number of dominant chromophores.

In order to compare the spectral information in the segmented nerve regions to the tissue environment, reference spectra were sampled from the ROIs, assuming a bivariate normal distribution $\mathcal{N}\left((\mu_{lat}, \mu_{ax}), \text{diag}(\sigma_{lat}^2, \sigma_{ax}^2)\right)$, where $\mu_{lat}, \mu_{ax}$ and $\sigma_{lat}, \sigma_{ax}$ are the pixelwise lateral and axial means and standard deviations of the locations of the considered nerves. These reference spectra are sampled for each of the nerves independently to account for the differences in the spectral environment due to fluence at different depths and due to the specifics of the surrounding anatomy.

C.3 Probabilistic data model

Based on the sparsity of the NMF, we propose a simple probabilistic model for the NMF coefficients that describes whether each spectral component is present in a tissue or not. This aspect is realized by a probabilistic mixture model with two components for the two corresponding situations where the respective coefficient is either zero or non-zero. In addition, as the distributions of the non-zero parts are strongly skewed, we performed Box-Cox power transformations to increase the comparability of different spectral components and the validity of metrics like the Pearson correlation coefficient. Since light fluence is thought to be a major contributor to skewness of the data, this transformation can also be seen as a partial fluence adjustment that improves the comparability of spectra at different depths.

In summary, we model the categorical parts of the mixture models for the nine spectral components via Bernoulli random variables $m_j \sim Bernoulli(p_j), j = 1, \dots, 9$, where $p_j$ is the probability that the corresponding coefficient is non-zero. On the level of the mixture components, the zero component is a deterministic variable $C_j^{(0)} \sim Det(0)$, while the non-zero component $C_j^{(1)}$ follows a distribution that is determined by the Box-Cox transformation of the non-zero coefficient data with power parameter $\beta_j$, which is determined via a maximum likelihood approach.

In order to compare the coefficients of different spectral components in an unbiased way (strong OA signal does not necessarily imply high clinical relevance), we introduced a standardization of the mixture models. Since the Box-Cox transformation results in distributions that are qualitatively similar to a normal distribution, we studentized the variables $C_j^{(1)}$ to obtain variables with zero mean and unit standard deviation $\tilde{C}_j^{(1)} \coloneqq (C_j^{(1)} - \mathbb{E}C_j^{(1)})/\sigma_j$, where $\sigma_j$ denotes the standard deviation of $C_j^{(1)}$. To keep the relation between very low values of these variables $\tilde{C}_j^{(1)}$ and the zero values, we shifted the deterministic components to the value $-3$ following the $3\sigma$ rule that states that for a normal distribution, 99.7% of the distribution lie within $3\sigma$. So, we introduce the standardized deterministic components $\tilde{C}_j^{(0)} \sim Det(-3)$, which are mixed with the components $\tilde{C}_j^{(1)}$ via the unchanged categorical variables $m_j$.

C.4 Correlation metrics

To compare the different spectral components on both levels of the model – the categorical and the continuous parts – we calculate correlation metrics. To see additional differences between nerve tissue and surrounding tissue, we performed this correlation analysis for the nerve data and the sampled reference data.

As correlation metric for the categorical variables we chose the pairwise Sørensen-Dice coefficient between the variables $m_j$,

$$DSC_{j,k} := \frac{2 m_j \cdot m_k}{|m_j|^2 + |m_k|^2} \in [0,1], \quad j,k = 1, \ldots, 9,$$

where the $m_j$ are interpreted as vectors, and $v \cdot w := \sum_j v_j w_j$ denotes the scalar product of two vectors. The Sørensen-Dice coefficient describes the amount of co-occurrences of two spectral components relative to the total occurrence of the two components. High values indicate that two spectral components usually appear together, while low values indicate that the spectral components rarely mix.

To study the pairwise correlation between the continuous variables $\tilde{C}_j^{(1)}$, we compute the Pearson correlation coefficients,

$$\rho_{j,k} := \frac{cov\left(\tilde{C}_j^{(1)}, \tilde{C}_k^{(1)}\right)}{\sigma_j \sigma_k} = \mathbb{E}\left[\tilde{C}_j^{(1)} \tilde{C}_k^{(1)}\right] \in [-1,1], \quad j,k = 1, \ldots, 9,$$

where $cov(X,Y) := \mathbb{E}[(X - \mathbb{E}X)(Y - \mathbb{E}Y)]$ is the covariance of two random variables $X$ and $Y$, $\sigma_j$ denotes the standard deviation of $\tilde{C}_j^{(1)}$ and the second equality follows from studentization. The Pearson correlation therefore describes how two spectral components mix when they mix: positive values for mixtures with a more or less fixed ratio, small values for random mixtures, and negative values for competitive mixtures.

C.5 Dimensionality reduction and hierarchical clustering

To visually access the full complexity of the spectral data set, and to investigate higher order correlations qualitatively, we embedded the 9-dimensional standardized spectral unmixing data into 2-dimensional space with the Uniform Manifold Approximation and Projection (UMAP) algorithm for dimensionality reduction (*36*).

The binary structure of our probabilistic model leads to a hierarchical structure in a natural way. Two spectra are in the same class if they are composed of the same spectral components. As a metric to compare the similarity of these classes, we propose using the Euclidean distance between the mean $L^2$-normalized spectra of the clusters, which is a metric for the similarity of the spectral shape. This approach links the class back to the original spectral data, making the clustering transparent and interpretable. For clustering linkage, we use Ward's linkage that minimizes in-cluster variance.

For our collection of nine spectral components, this approach leads to at most $2^9 = 512$ leaves of the clustering tree. To compare tissues based on this hierachical clustering, we determine the distribution of the spectral data of ROIs over the found clusters. We call this distribution the spectral fingerprint of a spectral dataset. In particular, this approach gives us a means to spectrally compare nerve tissue to the surrounding tissue.

## References

1. A. Tagliafico, B. Bignotti, C. Martinoli, Update on Ultrasound-Guided Interventional Procedures on Peripheral Nerves. *Semin Musculoskelet Radiol* **20**, 453-460 (2016).
2. F. Mirza, A. R. Brown, Ultrasound-guided regional anesthesia for procedures of the upper extremity. *Anesthesiol Res Pract* **2011**, 579824 (2011).
3. D. W. Frijlink, G. J. Brekelmans, L. H. Visser, Increased nerve vascularization detected by color Doppler sonography in patients with ulnar neuropathy at the elbow indicates axonal damage. *Muscle Nerve* **47**, 188-193 (2013).
4. N. L. Gonzalez, L. D. Hobson-Webb, Neuromuscular ultrasound in clinical practice: A review. *Clin Neurophysiol Pract* **4**, 148-163 (2019).
5. G. Rangavajla, N. Mokarram, N. Masoodzadehgan, S. B. Pai, R. V. Bellamkonda, Noninvasive imaging of peripheral nerves. *Cells Tissues Organs* **200**, 69-77 (2014).
6. S. B. Raval, C. A. Britton, T. Zhao, N. Krishnamurthy, T. Santini, V. S. Gorantla, T. S. Ibrahim, Ultra-high field upper extremity peripheral nerve and non-contrast enhanced vascular imaging. *PLoS One* **12**, e0175629 (2017).
7. G. Barisano, F. Sepehrband, S. Ma, K. Jann, R. Cabeen, D. J. Wang, A. W. Toga, M. Law, Clinical 7 T MRI: Are we there yet? A review about magnetic resonance imaging at ultra-high field. *Br J Radiol* **92**, 20180492 (2019).
8. S. Ibrahim, N. D. Harris, M. Radatz, F. Selmi, S. Rajbhandari, L. Brady, J. Jakubowski, J. D. Ward, A new minimally invasive technique to show nerve ischaemia in diabetic neuropathy. *Diabetologia* **42**, 737-742 (1999).
9. A. J. Boulton, Diabetic neuropathy: classification, measurement and treatment. *Curr Opin Endocrinol Diabetes Obes* **14**, 141-145 (2007).
10. N. E. Cameron, S. E. Eaton, M. A. Cotter, S. Tesfaye, Vascular factors and metabolic interactions in the pathogenesis of diabetic neuropathy. *Diabetologia* **44**, 1973-1988 (2001).
11. E. Suzuki, K. Yasuda, K. Yasuda, S. Miyazaki, N. Takeda, H. Inouye, N. Omawari, K. Miura, 1H-NMR analysis of nerve edema in the streptozotocin-induced diabetic rat. *J Lab Clin Med* **124**, 627-637 (1994).
12. N. P. Gonçalves, C. B. Vægter, H. Andersen, L. Østergaard, N. A. Calcutt, T. S. Jensen, Schwann cell interactions with axons and microvessels in diabetic neuropathy. *Nat Rev Neurol* **13**, 135-147 (2017).

13. R. Li, E. Phillips, P. Wang, C. J. Goergen, J. X. Cheng, Label-free in vivo imaging of peripheral nerve by multispectral photoacoustic tomography. *J Biophotonics* **9**, 124-128 (2016).
14. J. M. Mari, W. Xia, S. J. West, A. E. Desjardins, Interventional multispectral photoacoustic imaging with a clinical ultrasound probe for discriminating nerves and tendons: an ex vivo pilot study. *J Biomed Opt* **20**, 110503 (2015).
15. T. P. Matthews, C. Zhang, D. K. Yao, K. Maslov, L. V. Wang, Label-free photoacoustic microscopy of peripheral nerves. *J Biomed Opt* **19**, 16004 (2014).
16. W. Xia, S. West, D. Nikitichev, S. Ourselin, P. Beard, A. Desjardins, *Interventional multispectral photoacoustic imaging with a clinical linear array ultrasound probe for guiding nerve blocks*. SPIE BiOS (SPIE, 2016), vol. 9708.
17. X. L. Deán-Ben, E. Merčep, D. Razansky, Hybrid-array-based optoacoustic and ultrasound (OPUS) imaging of biological tissues. *Applied Physics Letters* **110**, 203703 (2017).
18. A. Taruttis, A. C. Timmermans, P. C. Wouters, M. Kacprowicz, G. M. van Dam, V. Ntziachristos, Optoacoustic Imaging of Human Vasculature: Feasibility by Using a Handheld Probe. *Radiology* **281**, 256-263 (2016).
19. J. Kukačka, S. Metz, C. Dehner, A. Muckenhuber, K. Paul-Yuan, A. Karlas, E. M. Fallenberg, E. Rummeny, D. Jüstel, V. Ntziachristos, Image processing improvements afford second-generation handheld optoacoustic imaging of breast cancer patients. *Photoacoustics* **26**, 100343 (2022).
20. A. Karlas, M. A. Pleitez, J. Aguirre, V. Ntziachristos, Optoacoustic imaging in endocrinology and metabolism. *Nat Rev Endocrinol* **17**, 323-335 (2021).
21. F. Knieling, C. Neufert, A. Hartmann, J. Claussen, A. Urich, C. Egger, M. Vetter, S. Fischer, L. Pfeifer, A. Hagel, C. Kielisch, R. S. Görtz, D. Wildner, M. Engel, J. Röther, W. Uter, J. Siebler, R. Atreya, W. Rascher, D. Strobel, M. F. Neurath, M. J. Waldner, Multispectral Optoacoustic Tomography for Assessment of Crohn's Disease Activity. *N Engl J Med* **376**, 1292-1294 (2017).
22. J. Reber, M. Willershäuser, A. Karlas, K. Paul-Yuan, G. Diot, D. Franz, T. Fromme, S. V. Ovsepian, N. Bézière, E. Dubikovskaya, D. C. Karampinos, C. Holzapfel, H. Hauner, M. Klingenspor, V. Ntziachristos, Non-invasive Measurement of Brown Fat Metabolism Based on Optoacoustic Imaging of Hemoglobin Gradients. *Cell Metab* **27**, 689-701.e684 (2018).
23. A. P. Regensburger, L. M. Fonteyne, J. Jüngert, A. L. Wagner, T. Gerhalter, A. M. Nagel, R. Heiss, F. Flenkenthaler, M. Qurashi, M. F. Neurath, N. Klymiuk, E. Kemter, T. Fröhlich, M. Uder, J. Woelfle, W. Rascher, R. Trollmann, E. Wolf, M. J. Waldner, F. Knieling, Detection of collagens by multispectral optoacoustic tomography as an imaging biomarker for Duchenne muscular dystrophy. *Nat Med* **25**, 1905-1915 (2019).
24. S. Tzoumas, A. Nunes, I. Olefir, S. Stangl, P. Symvoulidis, S. Glasl, C. Bayer, G. Multhoff, V. Ntziachristos, Eigenspectra optoacoustic tomography achieves quantitative blood oxygenation imaging deep in tissues. *Nat Commun* **7**, 12121 (2016).
25. J. Vonk, J. Kukačka, P. J. Steinkamp, J. G. de Wit, F. J. Voskuil, W. T. R. Hooghiemstra, M. Bader, D. Jüstel, V. Ntziachristos, G. M. van Dam, M. J. H. Witjes, Multispectral optoacoustic tomography for in vivo detection of lymph node metastases in oral cancer patients using an EGFR-targeted contrast agent and intrinsic tissue contrast: A proof-of-concept study. *Photoacoustics* **26**, 100362 (2022).
26. K. B. Chowdhury, J. Prakash, A. Karlas, D. Justel, V. Ntziachristos, A Synthetic Total Impulse Response Characterization Method for Correction of Hand-held Optoacoustic Images. *IEEE Trans Med Imaging*, (2020).
27. K. B. Chowdhury, M. Bader, C. Dehner, D. Jüstel, V. Ntziachristos, Individual transducer impulse response characterization method to improve image quality of array-based handheld optoacoustic tomography. *Opt. Lett.* **46**, 1-4 (2021).
28. C. Dehner, I. Olefir, K. B. Chowdhury, D. Justel, V. Ntziachristos, Deep-learning-based electrical noise removal enables high spectral optoacoustic contrast in deep tissue. *IEEE Trans Med Imaging* **Pp**, (2022).
29. A. Longo, J. D, V. Ntziachristos, Disentangling the frequency content in optoacoustics. *IEEE Transactions on Medical Imaging*, 1-1 (2022).

30. E. L. Feldman, B. C. Callaghan, R. Pop-Busui, D. W. Zochodne, D. E. Wright, D. L. Bennett, V. Bril, J. W. Russell, V. Viswanathan, Diabetic neuropathy. *Nat Rev Dis Primers* **5**, 42 (2019).
31. N. Akcar, S. Ozkan, O. Mehmetoglu, C. Calisir, B. Adapinar, Value of power Doppler and gray-scale US in the diagnosis of carpal tunnel syndrome: contribution of cross-sectional area just before the tunnel inlet as compared with the cross-sectional area at the tunnel. *Korean J Radiol* **11**, 632-639 (2010).
32. M. Masthoff, A. Helfen, J. Claussen, A. Karlas, N. A. Markwardt, V. Ntziachristos, M. Eisenblätter, M. Wildgruber, Use of Multispectral Optoacoustic Tomography to Diagnose Vascular Malformations. *JAMA Dermatol* **154**, 1457-1462 (2018).
33. M. Lorenzi, The polyol pathway as a mechanism for diabetic retinopathy: attractive, elusive, and resilient. *Exp Diabetes Res* **2007**, 61038 (2007).
34. R. Stotzka, N. Ruiter, T. Mueller, R. Liu, H. Gemmeke, *High resolution image reconstruction in ultrasound computer tomography using deconvolution*. Medical Imaging (SPIE, 2005), vol. 5750.
35. A. Cichocki, A.-H. Phan, Fast local algorithms for large scale nonnegative matrix and tensor factorizations. *IEICE transactions on fundamentals of electronics, communications and computer sciences* **92**, 708-721 (2009).
36. L. McInnes, J. Healy, J. Melville, UMAP: uniform manifold approximation and projection for dimension reduction. (2020).

**List of abbreviations**

| | |
|---|---|
| OPUS | hybrid optoacoustic and ultrasound |
| US | ultrasonography |
| MRI | magnetic resonance imaging |
| PET | positron emission tomography |
| OA | optoacoustic |
| MSOT | multispectral optoacoustic tomography |
| ROI | region of interest |
| HHb | deoxygenated hemoglobin |
| $HbO_2$ | oxygenated hemoglobin |
| DSC | Sørensen-Dice coefficient |
| PCC | Pearson correlation coefficient |
| UMAP | uniform manifold approximation and projection |
| BMI | body mass index |
| NMF | non-negative matrix factorization |

**Acknowledgments**

The authors would like to thank Robert Wilson, Gabriella Leung, and Sergey Sulima for proofreading and improving the manuscript and Magda Paschali for valuable discussions.

This project has received funding from the European Research Council (ERC) under the European Union's Horizon Europe research and innovation programme under grant agreement No 101041936 (EchoLux) and the Horizon 2020 research and innovation programme under grant agreement No 694968 (PREMSOT).

**Funding**

European Research Council (ERC), Horizon Europe research and innovation programme, grant 101041936 (DJ)

European Research Council (ERC), Horizon 2020 research and innovation programme, grant 694968 (VN)

**Author contributions**

Conceptualization: DJ, HI, WS, FH

Methodology: DJ, HI, FH, CD

Investigation: DJ, HI, WS, FH

Visualization: DJ

Funding acquisition: VN, GS, NN

Project administration: VN

Supervision: DJ, WS, NN, GS, VN

Writing – original draft: DJ, HI

Writing – review & editing: DJ, HI, FH, CD, WS, NN, GS, VN

**Competing Interest**

Vasilis Ntziachristos is an equity owner and consultant at iThera Medical GmbH.

**Data and materials availability**

The data that support the findings of this study are available on request from the corresponding author. The data are not publicly available due to privacy or ethical restrictions.

## Supplementary Materials

## Supplementary Figures

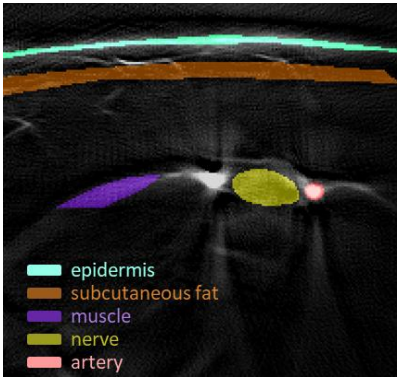

**Supplementary Figure 1. Segmentation of tissues.** The segmentations of the six tissues analysed in Fig. 2. To properly capture the spectral contrast, the effect of fluence attenuation was reduced by only segmenting the superficial layers of bulk tissues, like subcutaneous fat or muscle.

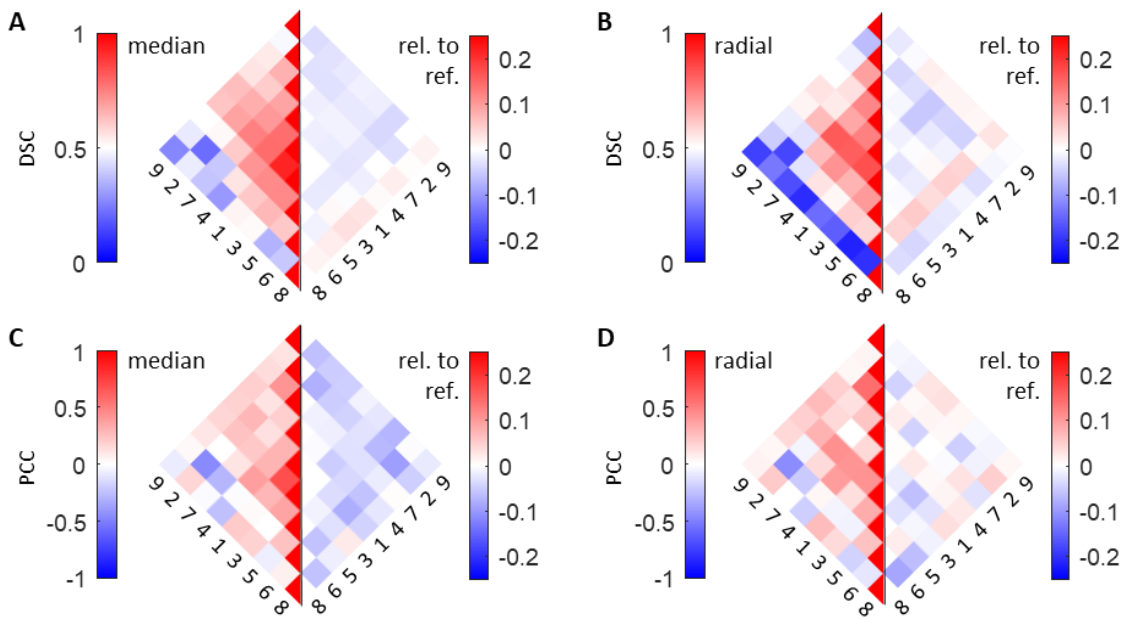

**Supplementary Figure 2. Correlation analysis for the spectral data acquired in median and radial nerves.** (**A**) and (**B**) Pairwise Sørensen-Dice coefficients (DSC) of the standardized spectral components of the median and radial nerve, respectively; and difference to the reference spectra that were sampled from the surrounding tissue. (**C**) and (**D**) Pairwise Pearson correlation coefficients (PCC) of the standardized continuous part of the spectral components of the median and radial nerve, respectively; and difference to the reference spectra that were sampled from the surrounding tissue.

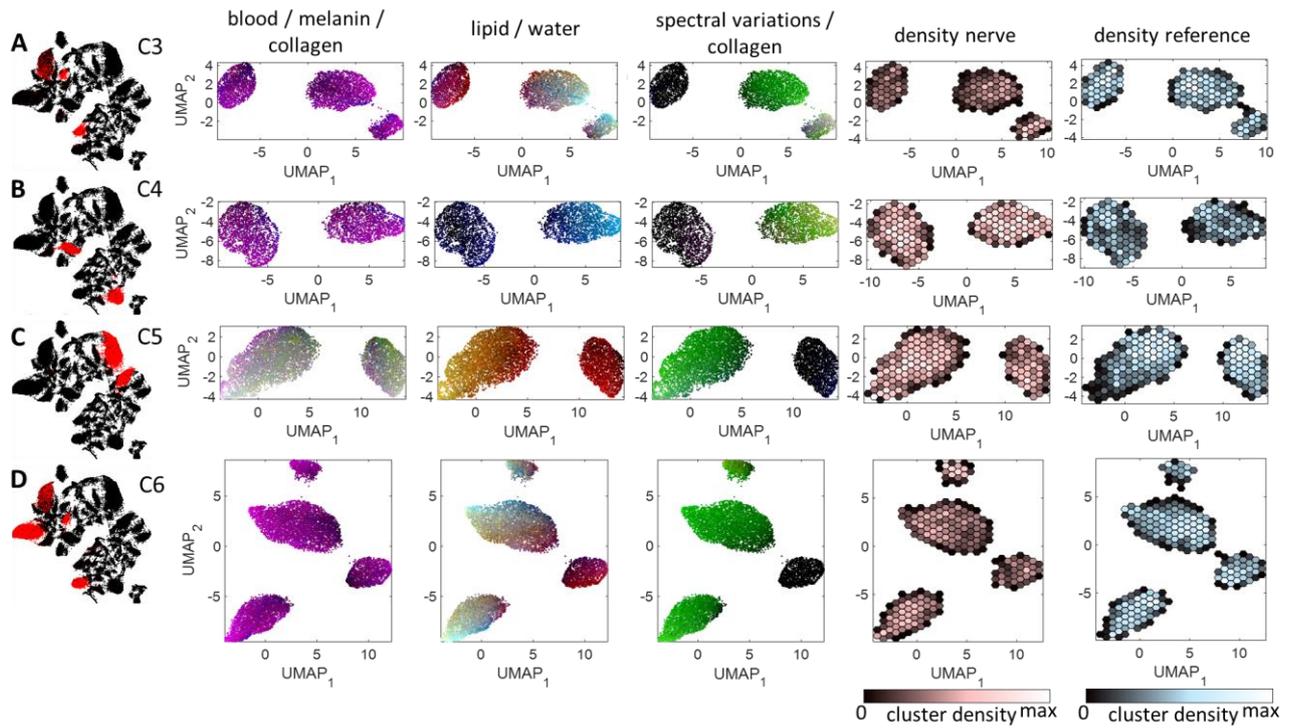

**Supplementary Figure 3. Details of the four clusters C3-C6 highlighted in Fig. 6D.** (**A**)-(**D**) The four clusters are localized in the full UMAP embedding (Fig. 6A-C) on the left. The UMAP embeddings of the clusters are color-coded as in Fig. 2 in the next three panels. Hexagonal density plots for the ulnar nerve data and the reference data are shown on the right.